\documentclass[twocolumn,showpacs,preprintnumbers,amsmath,
  amssymb,showkeys,floatfix,prl,dvips]{revtex4}

\usepackage{graphics}
\usepackage{graphicx}
\usepackage{dcolumn}
\usepackage{bm}
\usepackage{verbatim}

\newcommand{\unitv}[1]{\hat{\bm{#1}}}

\begin{document}

\title{Observation of a Turbulence--Induced Large Scale Magnetic
  Field}

\author{E.\;J.\;Spence}
\author{M.\;D.\;Nornberg}
\author{C.\;M.\;Jacobson} 
\author{R.\;D.\;Kendrick}
\author{C.\;B.\;Forest}
\email{cbforest@wisc.edu}
\affiliation{
  Department of Physics,  University of Wisconsin--Madison,
  1150 University Avenue, Madison, Wisconsin 53706
}

\date{\today}

\begin{abstract}
An axisymmetric magnetic field is applied to a spherical, turbulent
flow of liquid sodium.  An induced magnetic dipole moment is measured
which cannot be generated by the interaction of the axisymmetric mean
flow with the applied field, indicating the presence of a turbulent
electromotive force.  It is shown that the induced dipole moment
should vanish for any axisymmetric laminar flow.  Also observed is the
production of toroidal magnetic field from applied poloidal magnetic
field (the $\omega$--effect).  Its potential role in the production of
the induced dipole is discussed.
\end{abstract}


\maketitle

Many stars and planets generate their own nearly--axisymmetric
magnetic fields.  Understanding the mechanism by which these fields
are generated is a problem of fundamental importance to astrophysics.
These dynamos are sometimes modeled using two components: a process
which generates toroidal magnetic field from poloidal field and a
feedback mechanism which reinforces the poloidal
field~\cite{Parker_1955}.  The first process is easily modeled in an
axisymmetric system: toroidal differential rotation of a
highly--conducting fluid sweeps the pre-existing poloidal field in the
toroidal direction creating toroidal field.  This phenomenon, known as
the $\omega$--effect, is efficient at producing magnetic field and has
been observed
experimentally~\cite{Lehnert_1957,Odier_et_al_1998,Bourgoin_et_al_2002}.
The second ingredient to the model is more subtle, as toroidal
currents must be generated to reinforce the original axisymmetric
poloidal field.  Cowling's theorem~\cite{Cowling_1933} excludes the
possibility of an axisymmetric flow generating such currents so some
symmetry--breaking mechanism is required.

The usual mechanism invoked~\cite{Moffatt} is a turbulent
electromotive force (EMF),
$\cal{E}=\left<\tilde{\mathbf{v}}\times\tilde{\mathbf{b}}\right>$,
whereby small scale fluctuations in the velocity and magnetic fields
break the symmetry and interact coherently to generate the large scale
magnetic field.  This EMF is sometimes
expanded~\cite{Krause_and_Raedler} in terms of transport coefficients
about the mean magnetic field:
\mbox{$\cal{E}=\alpha\mathbf{B}+\beta\nabla\times\mathbf{B}+\bm{\gamma}\times\mathbf{B}$};
$\alpha$ is characterized by helicity in the turbulence, $\beta$ by
enhanced diffusion and $\bm{\gamma}$ by a gradient in the intensity of
the turbulence.  $\alpha$ is of particular interest as it results in
current flowing parallel to a magnetic field, and when coupled with
the $\omega$--effect can generate the toroidal currents needed to
reinforce the poloidal field.

\begin{figure}
\includegraphics[width=0.8\columnwidth]{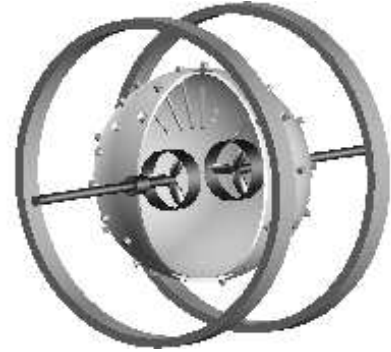}
\caption{Schematic of the Madison Dynamo Experiment showing a cut-away
view of the sphere, impellers, external field coils, surface and
internal Hall probes.}
\label{fig:schematic}
\end{figure}

Experimental evidence for mean--field EMFs (such as the
$\alpha$--effect) in turbulent flows has been scarce.  Three
experiments, relying on a laminar $\alpha$--effect, have generated an
EMF~\cite{Steenbeck_et_al_1968} and dynamo
action~\cite{Gailitis_et_al_2000,Stieglitz_and_Muller_2001}, but
heavily-constrained flow geometries were used to produce the needed
helicity; the role of turbulence was ambiguous.  Experiments with
unconstrained flows have provided evidence for turbulent EMFs, though
not the turbulent $\alpha$--effect.  Reighard and
Brown~\cite{Reighard_and_Brown_2001} have attributed a measured
reduction in the conductivity of a turbulent flow of sodium to the
$\beta$--effect.  P\'{e}tr\'{e}lis et al.\ have
observed~\cite{Petrelis_et_al_2003} distortion of a magnetic field
similar to an $\alpha$--effect (currents generated in the direction of
an applied magnetic field) and postulate that turbulence may be
responsible for disagreement between a laminar model and observations.
Not all liquid--metal experiments have had such results: Frick et al.\
have reported~\cite{Frick_et_al_2004} that the mean flow accounts for
all magnetic fields in their torus--shaped gallium experiment, and
Peffley, Cawthorne and
Lathrop~\cite{Peffley_and_Cawthorne_and_Lathrop_2000} have observed no
such effects.  It should also be noted that an $\alpha$--effect has
been observed in the core of magnetically--confined
plasmas~\cite{Ji_et_al_1996,Redd_et_al_2002}.

In this Letter we report measurements of the magnetic field induced by
applying an axisymmetric magnetic field to a turbulent, axisymmetric
flow of liquid sodium.  An induced dipole moment is measured which
cannot be generated by the mean flow, indicating the presence of a
turbulent EMF.

\begin{figure}
\includegraphics[width=0.9\columnwidth]{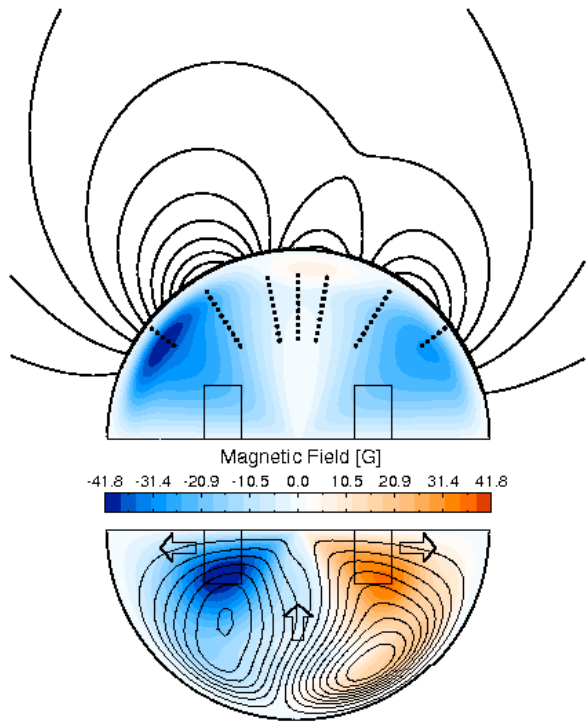}
\caption{{\it Upper half:} color contours of induced toroidal magnetic
  field, $B_\phi(s,z)$, measured by sets of internal Hall probes, for
  $Rm_{tip}=100$.  Induced poloidal flux surfaces, $\Psi(s,z)$, are in
  black.  The positions of the internal Hall probes are indicated with
  dots. The cylindrical axis of symmetry is horizontal.  {\it Lower
  half:} velocity field measured in a water model of the Madison
  Dynamo Experiment, for an impeller rotation rate of 16.7\,Hz.
  Contours of toroidal flow, $v_\phi(s,z)$, are in color and poloidal
  stream function, $\Phi(s,z)$, are in black.  The arrows indicate the
  direction of the poloidal flow, and the rectangles indicate the
  positions and size of the impellers which drive the flow.}
\label{fig:flows}
\end{figure}

The study is conducted in the Madison Dynamo Experiment, a 1\,m
diameter stainless steel sphere containing liquid sodium.  As shown in
Fig.~\ref{fig:schematic}, two drive shafts enter the sphere through
each pole and drive 30.5\,cm diameter impellers which generate an
axisymmetric mean flow.  The shafts are coupled to two 75\,kW motors
which are independently controlled by variable--frequency drives.  The
radial component of the magnetic field is measured by an array of 74
temperature--compensated Hall probes mounted to the sphere's surface,
allowing resolution of spherical harmonic components of the external
magnetic field up to polar order of $\ell=7$ and azimuthal order of
$m=5$.  Magnetic fields within the sphere are measured by seven linear
arrays of Hall probes inserted into the sodium within stainless steel
sheaths.  These probes are oriented to measure either the axial or
toroidal component of the field.  Finally, two external
electromagnets, in a Helmoltz configuration coaxial with the
impellers, apply a nearly uniform magnetic field throughout the
sphere. The applied field is between 0 and 60\,G, and dominated by
spherical harmonic content of $\ell=1,m=0$; the largest measured
$m\ne0$ component of the applied field is less than 2\% of the
axisymmetric part.

The study is conducted in the kinematic regime---the magnetic field is
not strong enough to affect the flow.  The strength of the Lorentz
force relative to the inertial forces acting on the fluid is
characterized by the interaction parameter (also called the Stuart
number), $N=\sigma a B_0^2 / \rho v_0$, where $a$ is the radius of the
sphere, $\sigma$ and $\rho$ are the conductivity and density of the
fluid, respectively, and $B_0$ and $v_0$ are characteristic magnetic
and velocity field magnitudes.  $N\sim10^{-2}$ for a total magnetic
field of 100\,G and $v_0=16.0$\,m/s, so the magnetic field is not
expected to alter the flow.  This is confirmed by the linear
dependence of the induced magnetic field with respect to the applied
field.  To affect the flow we would expect $N\approx 0.1$, or
$B_0\approx 180$\,G, a field magnitude not yet achieved.  We note that
the fluctuations, which are characterized by slower velocities, may be
in a regime that is affected by the magnetic field.

The axisymmetric part of the velocity field generated by the impellors
can be expressed in cylindrical coordinates $(s,\phi,z)$ as
\begin{equation}
\mathbf{v}=\nabla\Phi\times\nabla\phi+v_\phi(s,z)\unitv{\phi},
\label{eq:Veq}
\end{equation}
where $\Phi(s,z)$ is the poloidal stream function.  The flow consists
of two large cells, one in the northern and one in the southern
hemisphere.  An example of this flow, based on measurements made in a
water model of the sodium apparatus~\cite{Forest_et_al_2002}, can be
seen in the lower half of Fig.~\ref{fig:flows}.  The poloidal cells
flow inward at the equator and outward at the poles.  The two toroidal
cells flow in opposing directions.  The flow is similar to the $t2s2$
flow proposed by Dudley and James~\cite{Dudley_and_James_1989}; a flow
which is calculated to magnetically self-excite at sufficiently high
magnetic Reynolds number, $Rm=\mu_0\sigma a v_0$, where $\mu_0$ is the
vacuum magnetic permeability ($Rm_{tip}=\mu_0\sigma a v_{tip}$, where
$v_{tip}$ is the impeller tip speed).  This study is conducted below
the critical $Rm$ for self-excitation, as demonstrated by the lack of
observed growing magnetic fields.
The Reynolds number of the fluid is $Re\sim 10^7$; turbulent
fluctuations of the measured flow can be as large as 20\% of the mean,
depending on location.

Once the sphere is full of sodium the motors are started and a
constant magnetic field is applied to the sphere.  Hall probes sample
the magnetic field at 1\,kHz for 5 minutes; the applied field is then
subtracted from these data to determine the induced field.
Measurements of the induced field are presented in the upper half of
Fig.~\ref{fig:flows}.  The field is represented by a toroidal
component, $B_\phi(s,z)$, and poloidal flux function, $\Psi(s,z)$,
such that
\begin{equation}
\mathbf{B} = \nabla \Psi \times \nabla \phi+B_\phi(s,z)\unitv{\phi}.
\label{eq:Beq}
\end{equation}
The toroidal magnetic field, undetectable by probes outside the sphere
and orthogonal to the applied poloidal field, is measured within the
sphere by internal Hall probes, confirming the presence of the
$\omega$--effect.  The peak amplitude of the toroidal magnetic field
scales linearly with $Rm$, and can be larger than the magnitude of the
applied field.

The external induced poloidal magnetic field is decomposed into its
spherical harmonic components to reveal its spatial structure.  Since
the Hall probes on the sphere's surface lie outside regions containing
currents the magnetic field can be expressed as the gradient of a
scalar magnetic potential, $\mathbf{B}=-\nabla\Phi_m$, which solves
Laplace's equation.  In spherical coordinates the solution to the
potential, for the region excluding the origin, is well known:
$\Phi_m(r,\theta,\phi)=\sum_{\ell,m}D_{\ell,m}r^{-(\ell+1)}Y_\ell^m(\theta,\phi)$,
where $Y_\ell^m(\theta,\phi)$ is the spherical harmonic.  The
coefficients in the expansion, $D_{\ell,m}$, which fit the mean
induced field are calculated using singular value decomposition.  The
induced poloidal magnetic field is predominantly axisymmetric; the
largest components are given in Tab.~\ref{tab:spectral_content}.  The
dominant components with $\ell$ equal to 3 and 5 are expected due to
the structure of the applied field and mean flow; the large measured
dipole component is not expected, as it cannot be generated by the
axisymmetric mean flow, as will be shown below.

\setlength{\extrarowheight}{1pt} 
\newcolumntype{e}{D{,}{,}{-1}}

\begin{table}[h]
\begin{ruledtabular}
\begin{tabular}{@{}e@{}@{}dd@{}d}
\multicolumn{1}{c}{\mbox{Harmonic ($\ell$,\,$m$)}} &
\multicolumn{1}{c}{\mbox{Energy}} & 
\multicolumn{1}{c}{\mbox{$B_{r,max}$}} &
\multicolumn{1}{c}{\mbox{RMS\;Fluctuation}}\\ 
\hline
1,\,0\,({\rm dipole}) &  1.6\;{\rm erg} & 
11.4\,{\rm G} & 1.8\,{\rm G} \\
2,\,0 &  0.2 &  3.1 & 3.5 \\
3,\,0 &  0.5 &  13.6 & 2.4 \\
4,\,0 &  0.1 & 7.1 & 3.5 \\
5,\,0 &  0.4 &  18.3 & 3.5\\
1,\,1 &  0.0 &   0.8 & 7.8\\
2,\,1 &  0.0 & 0.6 & 3.3 \\
\end{tabular}
\end{ruledtabular}
\caption{Mean energy in the largest induced external poloidal
  harmonics, maximum mean radial field on the sphere's surface, and
  field fluctuation level for several spherical harmonic components,
  for $Rm_{tip}=100$ and an applied field of 60\,G.}
\label{tab:spectral_content}
\end{table}

\begin{figure}
\includegraphics[width=0.9\columnwidth]{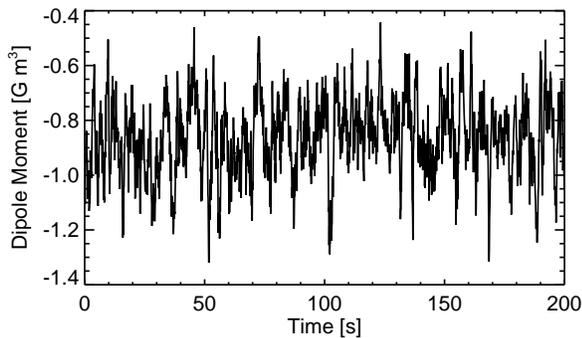}
\caption{Induced dipole moment versus time, for $Rm_{tip}=100$ and an
  applied magnetic field of 60\,G.  1\,G\,m$^3$ corresponds to 13.2\,G
  at the sphere's pole.}
\label{fig:S1signal}
\end{figure}

\begin{figure}[t]
\includegraphics[width=0.9\columnwidth]{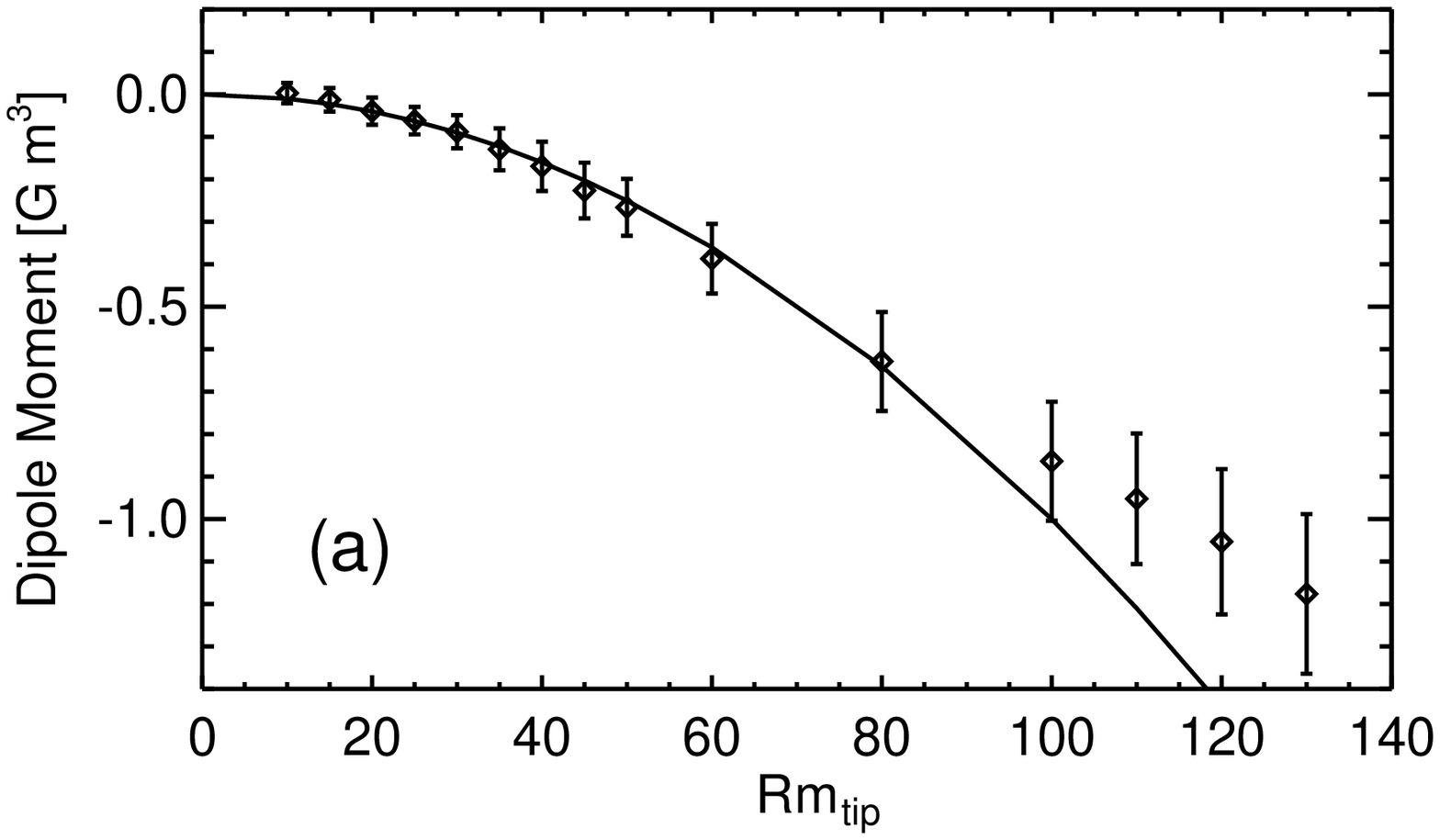}\\
\vspace{0.2cm}
\includegraphics[width=0.9\columnwidth]{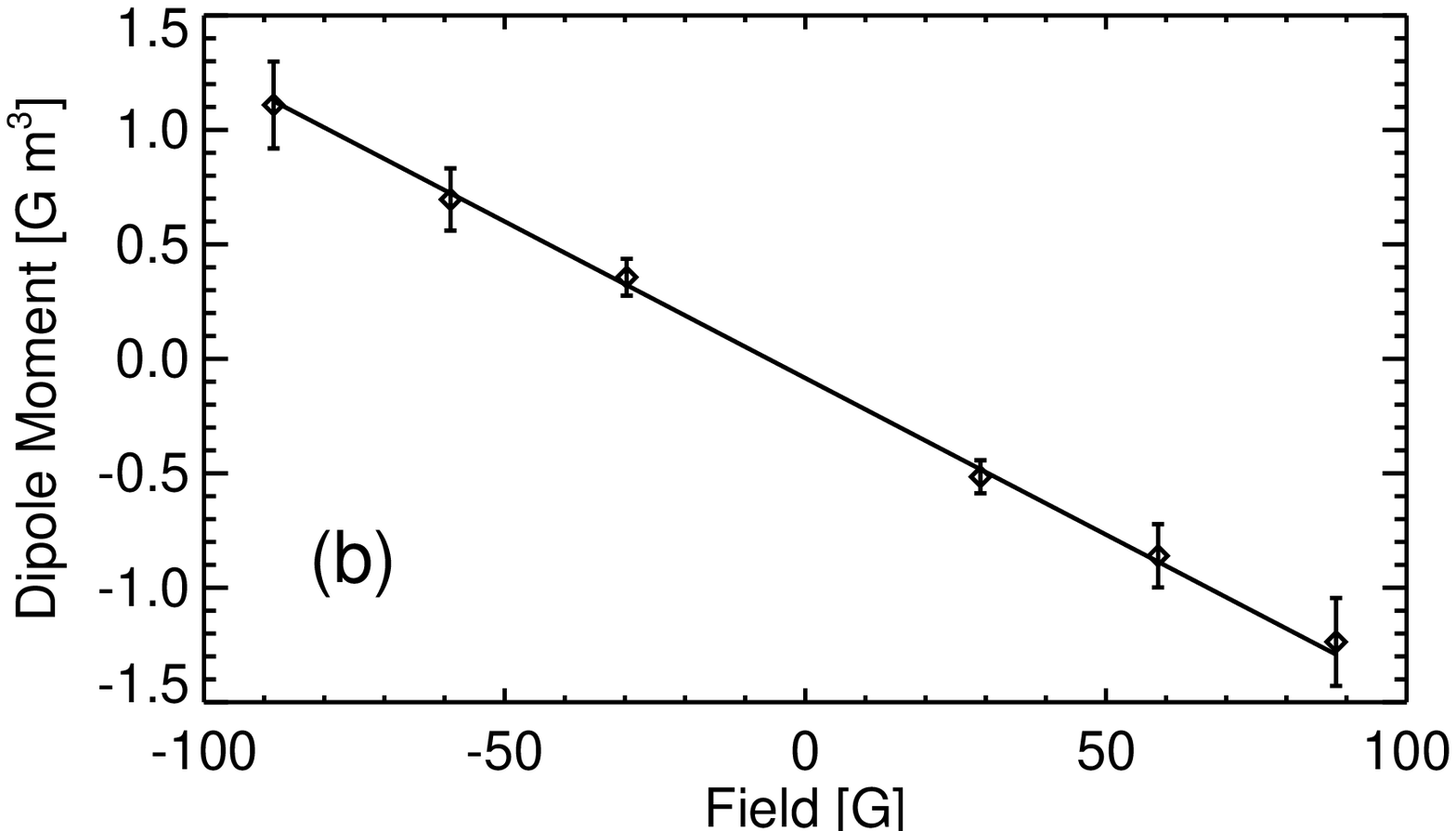}
\caption{ (a) Mean induced dipole moment versus $Rm_{tip}$, for an
  applied field of 60\,G with a quadratic fit valid at low $Rm$.  (b)
  Mean induced dipole moment versus applied magnetic field, for
  $Rm_{tip}=100$.  A linear fit is plotted for comparison.  Error bars
  are RMS fluctuation levels about the mean; the uncertainties in the
  mean values are very small (less than 0.01\,G\,m$^3$) due to long
  averaging times.}
\label{fig:Dipole_scaling}
\end{figure}

The induced dipole moment fluctuates dramatically in time around a
well--defined mean, as seen in Fig.~\ref{fig:S1signal}.  Measurements
indicate that the induced dipole depends on $Rm$
(Fig.~\ref{fig:Dipole_scaling}a) and upon the magnitude of the
externally--applied field (Fig.~\ref{fig:Dipole_scaling}b).  The
dipole moment's dependence on $Rm$ eliminates the possibility of the
measurement being a systematic error in the analysis.  The EMF depends
linearly on the applied field, indicating that it is a kinematic
effect and not due to the back reaction.

While Cowling's theorem demonstrates that self-excitation is not
possible in axisymmetric systems, it is not obvious that a dipole
moment cannot be induced by an axisymmetric velocity field exposed to
an axisymmetric magnetic field.  The proof of this is as follows.
Consider a bounded, steady--state, axisymmetric system described by
Eq.~\ref{eq:Beq}.  For axisymmetric fields, the only non-trivial
component of the dipole moment,
$\boldsymbol{\mu}\equiv\int\mathbf{x}\times\mathbf{J}\,d^3x$, is
oriented along the symmetry axis and results from currents flowing in
the toroidal direction,
\begin{equation}
\mu_z = \int s J_\phi\, d^3x.
\label{eq:muz}
\end{equation}
These currents can only be generated by the
$\mathbf{v}\times\mathbf{B}$ force due to the mean fields, so using
Ohm's law gives
\begin{eqnarray}\nonumber
s J_\phi & = & s\sigma\left[\mathbf{v}\times\left( \nabla \Psi \times
  \nabla \phi \right)\right]\cdot\unitv{\phi} 
\\ \nonumber & = &
\sigma\left[v_\phi\nabla\Psi-\left(\mathbf{v}\cdot\nabla\Psi\right)\unitv{\phi}\right]\cdot\unitv{\phi}
\\ & = &
-\sigma\nabla\cdot\left(\mathbf{v}\Psi\right),\label{eqn:ohm}
\end{eqnarray}
where use has been made of \mbox{$\nabla\Psi\cdot\unitv{\phi}=0$} and
the fluid has been assumed incompressible,
\mbox{$\nabla\cdot\mathbf{v}=0$}.  Inserting Eq.\ \ref{eqn:ohm} into
Eq.~\ref{eq:muz} and making use of Gauss' theorem and
$\mathbf{v}\cdot\unitv{n}=0$, where $\unitv{n}$ is the unit vector
normal to the vessel's surface, one finds that $\mu_z=0$.  It is
interesting to note that it is only the dipole moment that vanishes;
moments which include different powers of $s$ in Eq.~\ref{eq:muz} are
nonzero in general.  This conclusion is also independent of geometry;
any simply--connected axisymmetric system gives the same result.

It is possible that an induced dipole could be generated if mean
non-axisymmetric magnetic and velocity field modes interacted.  The
stainless steel tubes which contain the internal Hall probes could
potentially break the symmetry and create a mean non-axisymmetric
flow.  However, if this were the case one would expect higher--order
non-axisymmetric induced field components, which are not observed (see
Tab.~\ref{tab:spectral_content}).  The mean induced dipole moment is
present both with and without the tubes.

Since it cannot be generated by the mean flow, the dipole moment must
be the result of turbulence breaking the symmetry of the system,
likely a turbulent EMF of some form.  Any of the terms in the
mean--field expansion of the EMF have the potential to yield the
observed mean dipole moment.  A toroidal $\alpha$--effect could
produce large scale toroidal currents by interacting with the observed
$\omega$--effect.  The small scale helicity needed for the
$\alpha$--effect might come from either a turbulent cascade or be
produced directly by the impellers.  The $\beta$--effect leads to
turbulent modifications of the fluid
conductivity~\cite{Krause_and_Raedler}.  A nonuniform $\beta$--effect
could cause uneven distributions of currents to generate the dipole
moment.  A third possibility is the
$\gamma$--effect~\cite{Krause_and_Raedler}, which expels magnetic
field from regions of high--intensity turbulence, resulting in
diamagnetism.  The intensity of the turbulence varies with position,
so the $\beta$--effect and the $\gamma$--effect are both candidates to
explain the field.

Expanding the EMF in terms of the mean magnetic field may not be
appropriate, since the largest fluctuations in the magnetic field do
not satisfy the scale--separation and homogeneity requirements usually
imposed in the expansion of the mean--field EMF.  The largest
turbulent magnetic fluctuations are $m=1$.  Their Gaussian probability
distribution is centered at zero, consistent with a
passively--advected magnetic field in a turbulent cascade of velocity
fluctuations.  These $m=1$ fluctuations in $\mathbf{B}$ could, in
principle, interact with $m=1$ fluctuations in the flow and average to
give a net toroidal current.

In summary, a mean dipole moment is induced in the experiment which
cannot be produced by the mean flow.  The induced currents are of the
correct form to create a poloidal magnetic field, as required in the
$\alpha\omega$--dynamo model~\cite{Parker_1955}.  This is the first
observation of this effect in a laboratory experiment.  Explicit
characterization of the EMF is impossible without more detailed
knowledge of the form of the turbulence and direct measurement of the
fluctuating components of $\mathbf{v}$ and $\mathbf{B}$.  Future work
will be directed towards identifying the characteristics of the
fluctuations responsible for producing the dipole field.  We also note
that no saturation of the mechanism has yet been definitively
observed, as might be expected from numerical simulations and
theory~\cite{Gruzinov_and_Diamond_1994,Cattaneo_and_Hughes_1996}.
Future experiments with larger magnetic fields may provide insight
into the saturation mechanism.

We express our gratitude to A. Bayliss for helpful dialogue and
C. Parada for assistance with data acquisition.  CBF would like to
thank S. Prager and P. Terry for their continued support and useful
discussions.  This work is funded by the US Department of Energy, the
National Science Foundation, and David and Lucille Packard Foundation.


\end{document}